\documentclass{article}
\pdfoutput=1

\usepackage[preprint,nonatbib]{neurips_2023}

\usepackage[utf8]{inputenc} 
\usepackage[T1]{fontenc}    
\usepackage{hyperref}       
\usepackage{url}            
\usepackage{booktabs}       
\usepackage{amsfonts}       
\usepackage{nicefrac}       
\usepackage{microtype}      
\usepackage{xcolor}         

\usepackage{amsfonts} 
\usepackage{amsmath}
\usepackage{mathabx}
\usepackage{bbm}

\usepackage{bm}
\usepackage{wrapfig}
\usepackage{caption}
\usepackage{graphicx}
\usepackage{multirow}
\usepackage{colortbl}

\usepackage{subfigure}
\usepackage{enumitem}
\usepackage{listings}

\usepackage{xcolor}
\usepackage{algorithm}
\usepackage{algorithmicx}
\usepackage{algpseudocode}
\definecolor{codegreen}{rgb}{0,0.6,0}
\definecolor{codegray}{rgb}{0.5,0.5,0.5}
\definecolor{codepurple}{rgb}{0.58,0,0.82}
\definecolor{backcolour}{rgb}{0.96,0.96,0.96}

\lstdefinestyle{mystyle}{
    backgroundcolor=\color{backcolour},   
    commentstyle=\color{codegreen},
    keywordstyle=\color{magenta},
    numberstyle=\tiny\color{codegray},
    stringstyle=\color{codepurple},
    basicstyle=\ttfamily\footnotesize,
    breakatwhitespace=false,         
    breaklines=true,                 
    captionpos=b,                    
    keepspaces=true,                 
    numbers=left,                    
    numbersep=5pt,                  
    showspaces=false,                
    showstringspaces=false,
    showtabs=false,                  
    tabsize=2
}

\title{Efficiently Predicting Protein Stability Changes Upon Single-point Mutation with Large Language Models}

\author{%
  Yijie Zhang$^{1,3*}$, Zhangyang Gao$^{1,2*}$, Cheng Tan$^{1,2}$\thanks{Equal contribution.}, and Stan Z. Li$^{1}$\thanks{Corresponding author.} \\
  $^{1}$AI Lab, Research Center for Industries of the Future, Westlake University  \\
  $^{2}$Zhejiang University \\
  $^{3}$McGill University \\
  \texttt{yj.zhang@mail.mcgill.ca} \\
  \texttt{\{tancheng,gaozhangyang,stan.zq.li\}@westlake.edu.cn} \\
}

\begin{document}

\maketitle

\begin{abstract}
Predicting protein stability changes induced by single-point mutations has been a persistent challenge over the years, attracting immense interest from numerous researchers. The ability to precisely predict protein thermostability is pivotal for various subfields and applications in biochemistry, including drug development, protein evolution analysis, and enzyme synthesis. Despite the proposition of multiple methodologies aimed at addressing this issue, few approaches have successfully achieved optimal performance coupled with high computational efficiency.
Two principal hurdles contribute to the existing challenges in this domain. The first is the complexity of extracting and aggregating sufficiently representative features from proteins. The second refers to the limited availability of experimental data for protein mutation analysis, further complicating the comprehensive evaluation of model performance on unseen data samples.
With the advent of Large Language Models(LLM), such as the ESM models in protein research, profound interpretation of protein features is now accessibly aided by enormous training data. Therefore, LLMs are indeed to facilitate a wide range of protein research. In our study, we introduce an ESM-assisted efficient approach that integrates protein sequence and structural features to predict the thermostability changes in protein upon single-point mutations. Furthermore, we have curated a dataset meticulously designed to preclude data leakage, corresponding to two extensively employed test datasets, to facilitate a more equitable model comparison. 
\end{abstract}

\section{Introduction}

Proteins are crucial biomolecules in organisms \cite{karplus2005molecular}, essential for catalyzing metabolic reactions \cite{michelucci2013immune,gill1978adp}, DNA replication \cite{benkovic2001replisome}, responding to stimuli \cite{zanchi2008mechanical}, providing cellular structure \cite{rego2012nonlinear}, and transporting molecules \cite{zeuthen2010water}. Over the past decades, advancements in electron microscopy have allowed researchers to study protein structure and functions at the molecular level, leading to the emergence of Protein Engineering \cite{hu2022protein,carter2011introduction}. This field, as well as other approaches towards different biomolecules, focuses on discovering~\cite{gao2023vqpl,ulmer1983protein,muir1998expressed,lutz2012protein},designing\cite{gao2022pifold,tan2023global,nips2019_ingraham,jing2021learning,gao2022alphadesign,tan2022rfold, protein_sovler, fold2seq, tan2023hierarchical, gao2023knowledge}, and realizing molecules with specific functions or properties\cite{jin2018junction,tan2023target,gao2023co,zhang2022e3bind, tan2023cross,sacha2021molecule, somnath2021learning, gao2022semiretro,gao2023motifretro}, contributing significantly to biochemistry and medicine.
Since any change in the components of protein molecules could be considered a chemical reaction, the thermostability could be interpreted as the tendency that which a mutation is about to happen, usually denoted as the subtraction of Gibbs folding free energy $\Delta G$ for a pair of wild and mutated proteins \ref{eq:ddg} \cite{tokuriki2009stability}. A  higher $\Delta\Delta G$ value indicates a more stable protein \cite{pey2007predicted}.
\begin{equation}
\label{eq:ddg}
    \Delta\Delta G_{MW} = \Delta G_M - \Delta G_W 
\end{equation}
where $M$ denotes mutated type and $W$ denotes wild type.

Previous researchers have proposed several approaches to predict the $\Delta\Delta G$ by exploring molecular dynamics to simulate protein folding patterns \cite{kollman2000calculating,pitera2000exhaustive}, analyzing statistical relationships on the mutation pairs \cite{thomas1996statistical,carter2001four,topham1997prediction,gilis1999prediction}, and utilizing molecular modeling techniques \cite{bordner2004large,guerois2002predicting}. The emergence of machine learning has brought this field ample possibilities \cite{capriotti2004neural,cao2022survey}. In general, machine learning-based protein thermostability prediction could be diversified into two categories. (i) Sequence-based methods mainly aim to extract useful features from protein sequences \cite{cheng2006prediction,huang2007sequence,fariselli2015inps,folkman2016ease,umerenkov2022prostata,montanucci2019ddgun,pancotti2021deep,li2021saafec}. (ii) Structure-based methods aim to extract the structural variances to predict the $\Delta\Delta G$ \cite{chen2020istable,pires2014duet,chen2020premps,li2020predicting,rodrigues2018dynamut,savojardo2016inps,buss2018foldx,dehouck2011popmusic,laimer2015maestro,diaz2023stability,barlow2018flex,wang2023pros,samaga2021scones}. Notably, PROSTATA \cite{umerenkov2022prostata} fine-tuned pretrained protein large language model to predict protein mutational effects and achieved considerable results.

In reviewing prior methodologies in this domain, two significant challenges emerge. First, achieving a standardized comparison among models remains problematic, even with the recent introduction of major testing datasets. This issue is exacerbated by many methods relying on private training datasets, some of which are not publicly accessible, hindering comprehensive comparisons. Moreover, as delineated in Appendix \ref{app:dataset}, certain public datasets exhibit information leakage when compared with common testing datasets. The second challenge lies in maintaining an equilibrium between model accuracy, computational demands, and reproducibility. For instance, while structure-based methods may outperform their sequence-based counterparts, they often come with substantial computational costs, especially when utilizing advanced protein structure predictors like AlphaFold2 \cite{schaap2001rosetta} and Rosetta \cite{jumper2021highly} to generate tertiary structures.


In this work, we present two primary contributions to address these challenges. Firstly, we constructed a cleaned protein single-point dataset containing protein wild and mutated sequences aligned with the backbone atom coordinates extracted from the PDB file, and the corresponding $\Delta \Delta G$ values. The dataset has two subsets, each filtering out similar structures corresponding to the S669 and Ssym datasets. Hence, researchers can now integrate experimental sequential and structural features, while also ensuring fair comparisons using the filtered training and test datasets, thereby eliminating information leakage. Secondly, we proposed an ESM-augmented model to predict the $\Delta \Delta G$ upon protein single-point mutation accurately and effectively. The source code and the datasets will be available publicly soon.

\section{Method}

\subsection{Dataset}

In this paper, we introduce a novel training dataset characterized by two principal contributions. First, we meticulously align sequences and structures, the latter representing the coordinates of the backbone atom within each amino acid. This alignment results in clean and precise sequence-structure pairs, ready for further applications. Second, we ensure no information leakage between our dataset and the widely recognized test datasets, specifically, the S669 \cite{pancotti2022predicting} and Ssym datasets \cite{pucci2018quantification}. Any structures bearing resemblance to those within the mentioned datasets have been diligently omitted to maintain integrity and uniqueness in our collection. Notably, we proposed two subsets corresponding to each test dataset to ensure not lose too many samples for training.

We derived our initial dataset from PROSTATA \cite{umerenkov2022prostata}, an aggregation of various datasets that have been utilized in preceding methodologies. This aggregation allowed for the selection of samples under unified experimental conditions. The dataset we propose encompasses both wild and mutated sequences, along with corresponding $\Delta \Delta G$ values, comprising $10544$ samples, including reverse mutations.
To preclude any potential information leakage between the training and test datasets, we employed the TM-Score, a measure signifying structural similarity amongst proteins, and analyzed it between each sample in both datasets. We adhered to the threshold and set it as an average score of 0.5~\cite{zhang2005tm}. Consequently, samples exhibiting a score exceeding this threshold, in comparison to any structures within the testing dataset, were excluded from our dataset. We have provided a comparative analysis in the Appendix \ref{app:dataset} to facilitate our structure selection.

Upon acquiring the refined training dataset, we extracted the backbone coordinates from the corresponding PDB files and aligned them with the sequences. This alignment aimed to generate coherent structure-sequence pairs, streamlining the feature extraction process. For the purposes of our experiment, we adopted the coordinates of the $C_{\alpha}$ atom to represent the structure of each residue.
However, discrepancies in alignment were observed during this process. To address these misalignments, we devised a checking algorithm to discern and select coherent and applicable sequence-structure pairs for subsequent experimentation. The specifics of this algorithm are delineated in the Appendix \ref{app:training_dataset}. Finally, we maintained two subsets for training purposes, with regards to evaluating the model performance on S669 or Ssym datasets.

\subsection{Model Details}

We separate the process of predicting \( \Delta \Delta G \) for single-point mutations into two steps. As per Equation ~\ref{eq:ddg}, we can predict the \( \Delta G \) value for each protein and then find the difference between these predictions to get \( \Delta \Delta G \). Building on the approach from CDConv~\cite{fan2022continuous}, we have integrated both sequence and structural information to improve our predictions. Given the lack of mutated structures, we have used the structures of wild proteins as our structural inputs, an approach supported by SCONES~\cite{samaga2021scones}, showing minimal structural changes in unstable protein mutations.

In particular, we used CDConv~\cite{fan2022continuous} for predicting the \( \Delta G \) value due to its innovative approach and well-structured graph convolutional network. Also, we have added the sequential embedding from the ESM-2 650M UR50D model to our input, enhancing our feature extraction capabilities. More details on the proposed strategies are provided below:

\begin{figure}[ht]
  \centering
  \centerline{\includegraphics[width=1\textwidth]{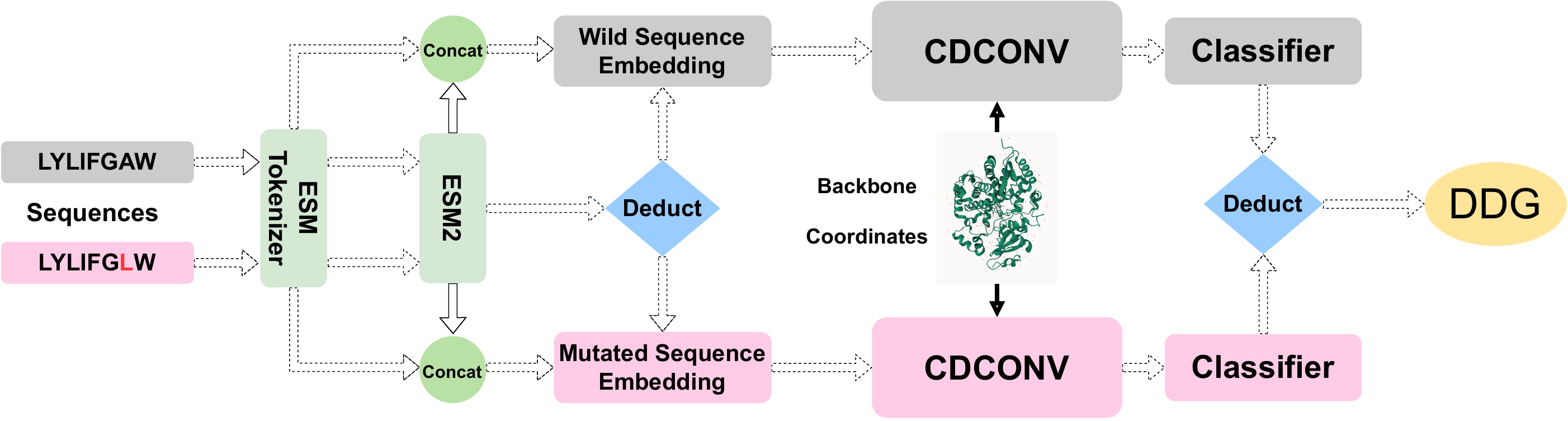}}
  \caption{The architecture of our proposed model}
  \label{fig:pipeline}
\end{figure}

\paragraph{System architeture}

Our model receives sequence-structure pairs for both wild-type and mutated-type as input. The ESM-assisted Continuous Graph Convolutional Network is independently employed on each input to assimilate sequential and structural features into a coherent embedding. Subsequently, we deduct one embedding from the other, transforming it into a one-dimensional prediction via a nonlinear classifier. The detailed system pipeline is shown in Figure \ref{fig:pipeline}.

\paragraph{ESM augmented variance enhancement}

To streamline the extraction of sequence features, we employed and froze the ESM-2 model to derive a sequential representation matrix for each mutation pair. Subsequently, the obtained embedding was concatenated with the sequence embedding, serving as the input for the graph neural networks. Given the scarce feature variances induced by single-point protein mutations, we also incorporated the difference between the mutated and wild sequence embeddings into both inputs to enhance the feature variance. We have presented the sample input generation process for either mutated or wild type in Algorithm \ref{algo:seqinput}. Consequently, this integration aids in accentuating the subtle alterations caused by mutations, allowing the model to effectively discern and learn from the nuanced differences between the mutated and wild-type sequences.
\begin{algorithm}
\caption{Model Input Generation}
\label{algo:seqinput}
\begin{algorithmic}[0]
\State $\textit{SeqInput} \gets \text{Concat}(\textit{SeqEmbed}_{\textit{ESM}}, \textit{SeqEmbed}_{\textit{CDConv}})$ 
\State $\textit{DiffEmbed} \gets \textit{SeqInput}_{\textit{mut}} - \textit{SeqInput}_{\textit{wild}}$ 
\State $\textit{FinalSeqInput} \gets \text{Concat}(\textit{SeqInput}, \textit{DiffEmbed})$ 
\end{algorithmic}
\end{algorithm}

\paragraph{Loss function}

In Appendix \ref{app:distribution}, the distribution of $\Delta \Delta G$ values reveals that the vast majority of samples are at a range between $[-5,5]$. Therefore, in order to better conceive the effect of major cases, we adopted the concept of Huber Loss, as denoted in Equation \ref{eq:huber}

\begin{equation}
\label{eq:huber}
L_{\delta}(y, f(x)) = 
\begin{cases} 
\frac{1}{2}(y - f(x))^2 & \text{for } |y - f(x)| \leq \delta \\
\delta |y - f(x)| - \frac{1}{2}\delta^2 & \text{otherwise}
\end{cases}
\end{equation}
Where \( L_{\delta} \) represents the Huber Loss, \( y \) is the true value, \( f(x) \) is the predicted value, and \( \delta \) is a threshold set to $1.0$.

\section{Experiment Details}

\paragraph{Datasets} 
We assess the effectiveness of our method utilizing two distinct test datasets: the S669 dataset \cite{pancotti2022predicting} and the Ssym dataset \cite{pucci2018quantification}. The details of curating the training dataset are illustrated in Appendix \ref{app:training_dataset}.

\begin{itemize}
    \item \textbf{S669 Dataset} is a novel test dataset that is less similar to conventional training sets, including 669 mutations. In order to better analyze the pattern of protein point mutation, we compiled S669\_r dataset, which is a reverse in $\Delta \Delta G$ value and the mutation points, based on the assumption that such reaction is reversible \cite{vila2022proteins}. For better evaluation, we employed a refined training dataset, encompassing $6951$ samples. These samples exhibit no structural similarity to those within the S669 dataset, ensuring the robustness and reliability of our assessment. 
    \item \textbf{Ssym Dataset} consists of an equal distribution of stabilizing and destabilizing mutations, each having 342 samples. Consequently, we have partitioned the dataset into Ssym and Ssym\_r subsets, including allocated stabilizing and destabilizing mutations following previous approaches \cite{umerenkov2022prostata, pucci2018quantification}, respectively. To evaluate our model performance on these datasets, we utilize an analogous cleaned training dataset containing $6080$ samples. These training samples are subjected to have no information leakage to the Ssym datasets, maintaining consistency in the evaluation process.
\end{itemize} 

\paragraph{Metrics} We chose to employ the Pearson Coefficient (PCC), Root Mean Square Error (RMSE), and Mean Absolute Error (MAE) as metrics to provide a comprehensive assessment of our model's performance. The Pearson Coefficient offers insights into the linear correlation between predicted and true values, RMSE accentuates larger errors providing a focus on substantial inaccuracies, and MAE presents a clear picture of the model’s predictive capability by measuring the average magnitude of the errors.

\paragraph{Baseline Models} 
We mainly utilized PROSTATA~\cite{umerenkov2022prostata} for comparison. We trained the model following the source code on two conditions. (i) We trained with the original sequence dataset that only excludes similar samples from each of the test datasets. For the S669, our filtered training dataset contains $7137$ samples, for the Ssym, it contains $6278$. (ii) We trained with our proposed dataset, which extracts the filtered sequential information from the structure-sequence pairs. For the S669, the training dataset has $6951$ samples while for the Ssym, it has $6080$. We reported the results and compared them with our model in Section \ref{exp:eval}. Besides, we compared our model's performance with previous works in Appendix \ref{app:add_val}.

\subsection{Evaluation on S669 and Ssym datasets}
\label{exp:eval}

Using our proposed dataset, we trained the model using two training datasets separately. One is for evaluating the model performance on the S669 and S669\_r datasets, and the other is for the Ssym and Ssym\_r datasets. In order to promote a fair utilization of all the data samples, we used a 5-fold cross-validation process for model training. The experiments were conducted under multiple hyperparameter combinations. In this way, we found the best hyperparameters for our model, illustrated in Appendix \ref{app:imple}. Then, we evaluated the performance of the model on the testing dataset by testing the model performance in each fold and averaging the 5 results. The results from both the original and reversed datasets are jointly presented in Tables \ref{tab:eval_669} and \ref{tab:eval:sym}, providing a more comprehensive assessment.
Together with the results of model performance under two testing datasets, we evaluated the average per-epoch training time of PROSTATA and our model under the same conditions. 

\begin{table}[ht]
\vspace{-2mm}
\centering
\caption{The evaluation results on the S669 and S669\_r datasets. }
\label{tab:eval_669}
\setlength{\tabcolsep}{1.5mm}{
\begin{tabular}{c|ccc|ccc|c}
\toprule
Method(Samples) & \multicolumn{3}{c|}{S669} & \multicolumn{3}{c}{S669\_r} & Runtime(minutes) \\
\cmidrule{2-4} \cmidrule{5-7}
& PCC  $\uparrow$ & RMSE  $\downarrow$ & MAE $\downarrow$ & PCC $\uparrow$ & RMSE $\downarrow$ & MAE $\downarrow$ \\
\midrule
PROSTATA(7137)          & 0.48 & 1.46 & 1.02 & 0.47 & 1.45 & 1.00 & - \\
PROSTATA(6951)          & 0.47 & 1.51 & 1.02 & 0.46 & 1.47 & 1.00 &  28.2(14.85 $\times$)\\
\hline
\rowcolor{gray!15} Ours(6951)                 & 0.51 & 1.46 & 1.07 & 0.51 & 1.46 & 1.07 & 1.9(1.00 $\times$)\\
\bottomrule
\end{tabular}}
\end{table}

\begin{table}[ht]
\vspace{-2mm}
\centering
\caption{The evaluation results on the Ssym and Ssym\_r datasets.}
\label{tab:eval:sym}
\setlength{\tabcolsep}{1.5mm}{
\begin{tabular}{c|ccc|ccc|c}
\toprule
Method(Samples) & \multicolumn{3}{c|}{Ssym} & \multicolumn{3}{c}{Ssym\_r} & Runtime(minutes)\\
\cmidrule{2-4} \cmidrule{5-7}
& PCC $\uparrow$ & RMSE  $\downarrow$ & MAE  $\downarrow$ & PCC  $\uparrow$ & RMSE  $\downarrow$ & MAE  $\downarrow$ \\
\midrule
PROSTATA(6278)          & 0.47 & 1.42 & 1.01 & 0.48 & 1.42 & 1.01 & - \\
PROSTATA(6080)          & 0.50 & 1.45 & 1.02 & 0.46 & 1.51 & 1.07 &23.9(14.94 $\times$)\\
\hline
\rowcolor{gray!15} Ours(6080)                & 0.47 & 1.45 & 1.05 & 0.47 & 1.45 & 1.05 & 1.6(1.00 $\times$)\\
\bottomrule
\end{tabular}}
\end{table}

From the results revealed, it could be concluded that for both testing datasets, the PROSTATA method with reduced samples shows a slight increase in RMSE and MAE compared to its counterpart, indicating a minor compromise in prediction accuracy.  Notably, our proposed method consistently performs on par with the PROSTATA method with similar sample sizes, while drastically improving the runtime efficiency, since our method freezes the gradient of the ESM model thus it does not need to update the parameters in it. In both testing datasets, our method achieves nearly 15 times faster processing than PROSTATA without compromising much on the accuracy metrics. These results highlight the efficacy of our method in balancing accuracy and computational efficiency.

\subsection{Ablation Study}

To elucidate the contribution of each component of our model, we performed a detailed ablation study. From Table \ref{tab:ablation}, our proposed model demonstrates the best performance across both datasets in all the metrics. When we remove the ESM embedding, there is a drastic decline in performance, especially in PCC values, which drop to 0.25 and 0.31 for S669 and Ssym, respectively. This suggests that ESM embedding plays a pivotal role in capturing the intricate patterns in the data. On the other hand, excluding the CDConv embedding results in a more subtle performance degradation. Although the decline is not as pronounced as in the ESM ablation, it is still noteworthy, particularly in the PCC metric. This indicates that while CDConv embedding contributes to the model's performance, the ESM embedding appears to be more critical for the model's efficacy.

\begin{table}[ht]
\centering
\caption{The ablation study of our model on two datasets.}
\setlength{\tabcolsep}{2.2mm}{
\begin{tabular}{ccccccccccc}
\toprule
\multirow{2}{*}{Method}    & \multicolumn{6}{c}{Metrics  } \\
& \multicolumn{2}{c}{PCC $\uparrow$} & \multicolumn{2}{c}{RMSE $\downarrow$} & \multicolumn{2}{c}{MAE $\downarrow$} \\
& S669 & Ssym & S669 & Ssym & S669 & Ssym \\
\midrule
Ours & 0.51 & 0.47 & 1.46 & 1.45 & 1.07 & 1.05 \\
w/o ESM embedding & 0.25 & 0.31 & 1.90 & 1.80 & 1.44 & 1.32 \\
w/o CDConv embedding & 0.48 & 0.39 & 1.54 & 1.54 & 1.13 & 1.12 \\
\bottomrule     
\end{tabular}}            
\label{tab:ablation}
\end{table}

\section{Conclusion and Discussion}
In our paper, we conducted thorough research on predicting the $\Delta \Delta G$ of protein single-point mutation. We assembled a dataset of aligned sequence-structure pairs, ensuring no information leakage with prevalent testing datasets to maintain fairness in model comparison and evaluation. Additionally, we introduced an ESM-augmented prediction model, leveraging both sequence and structure features, demonstrating exemplary performance across all testing datasets.

In our approach, the ESM model is frozen during the acquisition of input embeddings, which are subsequently stored locally, enhancing the model's computational efficiency. This approach relies solely on authentic PDB files, avoiding the use of computational protein structure predictors like AlphaFold2, making it more accessible for researchers to conduct experiments independently.

However, our approach has its limitations. The refinement of sequence selection algorithms is crucial during the cleaning of sequences with structural features, and further investigations into biochemical aspects are necessary to improve the robustness and interoperability of the model.

\bibliography{main}
\bibliographystyle{unsrt}

\clearpage
\newpage

\appendix
\section{Dataset details}
\label{app:dataset}

\subsection{Training dataset}
\label{app:training_dataset}
\begin{figure}[ht]
\centering
\includegraphics[width=1\textwidth]{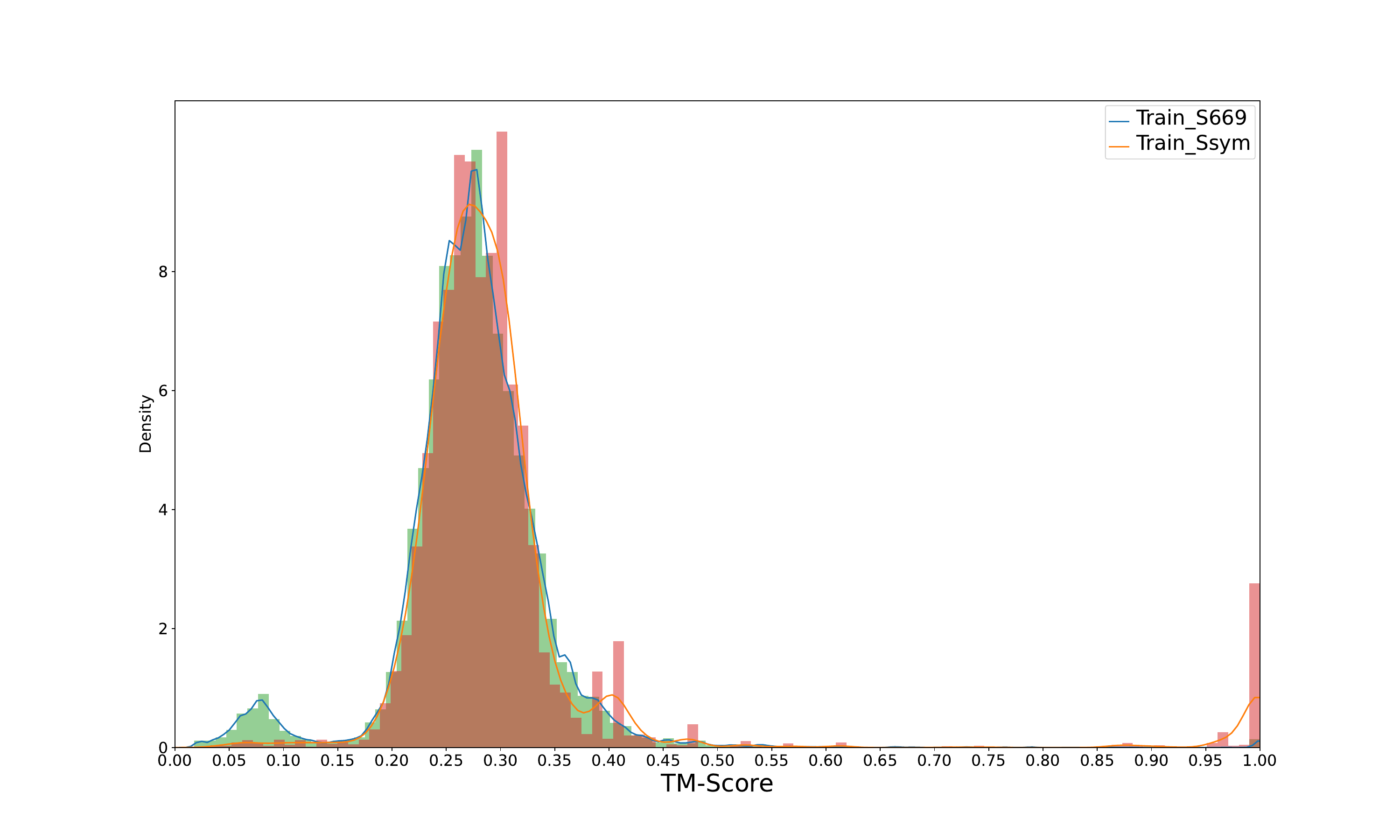}
\caption{Density distribution of the original dataset towards testing datasets} 
\label{fig:app:ori_distri}
\end{figure}

\begin{figure}[ht]
\centering
\includegraphics[width=1\textwidth]{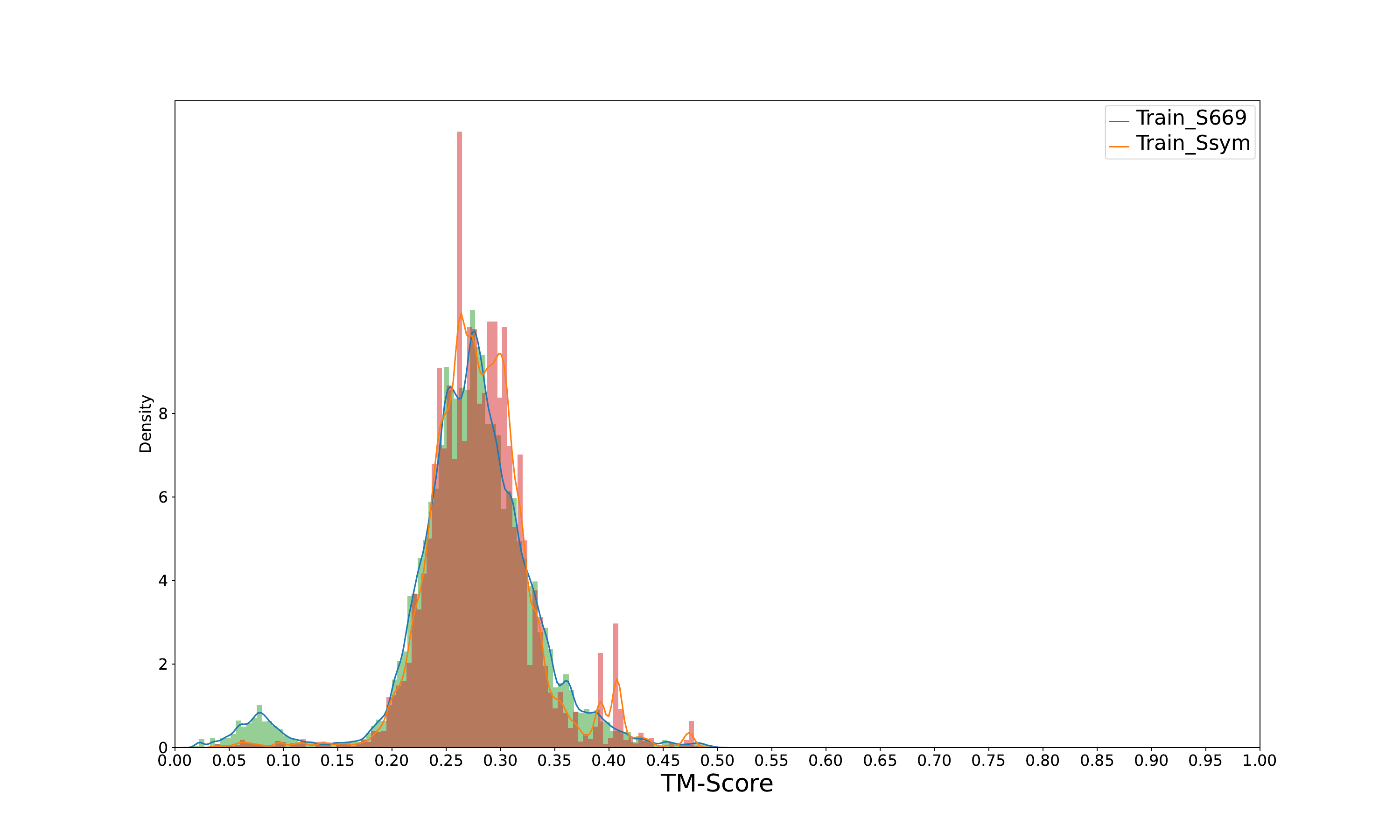}
\caption{Density distribution of our proposed datasets towards testing datasets} 
\label{fig:app:mut_distri}
\end{figure}

We collected samples from PROSTATA ~\cite{umerenkov2022prostata}, subsequently executing TM-Score calculations between each sample pair in the original dataset and those within the S669 and Ssym datasets, intended for evaluative comparisons. Subsequently, samples from the original dataset exhibiting a TM-Score exceeding 0.5 with any sample in the testing datasets were deliberately omitted since such a score indicates a notable structural similarity. This task was performed separately for each testing dataset, ensuring more equitable comparisons without compromising the richness of useful samples. We’ve depicted the structural similarity distribution between the original dataset and the test datasets in Figure \ref{fig:app:ori_distri}, and the distribution after the omission of similar samples is illustrated in Figure \ref{fig:app:mut_distri}. The original dataset contains $10544$ samples, after filtering similar structures, the subset for testing the S669 dataset has $6951$ sample while the subset for the Ssym dataset has $6080$. From this, it is evident that the original dataset revealed a substantial number of similar samples, which have been excluded from the filtered datasets. Also, it demonstrates that there are significantly more similar samples between the Ssym dataset and the original training datasets.

\subsection{Data Cleaning}

\begin{figure}[ht]
\centering
\includegraphics[width=0.8\textwidth]{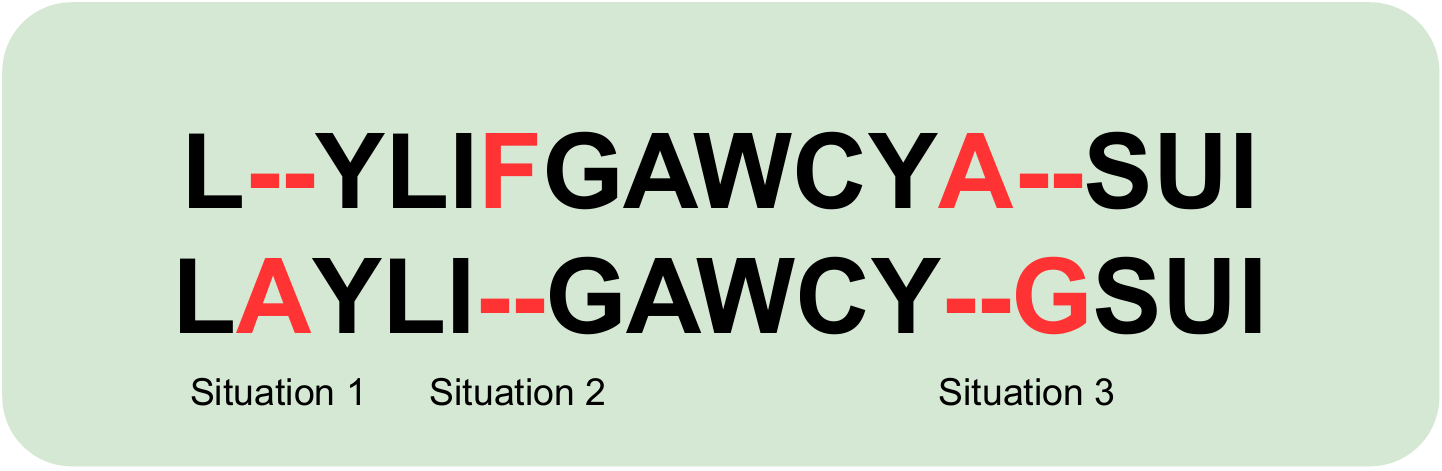}
\caption{The situations considered in the data cleaning process} 
\label{fig:app:check}
\end{figure}

We first aligned the original sequences and mutated sequences separately with the sequences extracted from PDB file using the Needleman-Wunsch algorithm. However, inconsistency often appears. In order to make strict alignment, we designed a cleaning procedure that deals with the following situations in the aligned sequences. The situations are simplified and visualized in Figure \ref{fig:app:check}. 

\begin{itemize}
    \item The provided sequence has missing residues. As situation 1 in Figrue \ref{fig:app:check} denotes, we amend the position with the corresponding residue in the PDB extracted sequence.
    \item The PDB extracted sequence has missing residues. As situation 2 in Figure \ref{fig:app:check} denotes, we following the PDB extracted sequence has the standard and thus omitted the residue from the sequence provided.
    \item The mutation normally appears as situation 3 denotes in Figure \ref{fig:app:check}. Under these circumstances, we reserve the mutated residue from the provided sequence.
\end{itemize}

It is worth noting that when the neighborhood residue of the mutated position is missing in PDB extracted structures, we are unable to tell if the mutant type is single-point or multiple-point. Therefore, we omitted the sequences under certain circumstances.

\begin{figure}[ht]
\centering
\includegraphics[width=0.8\textwidth]{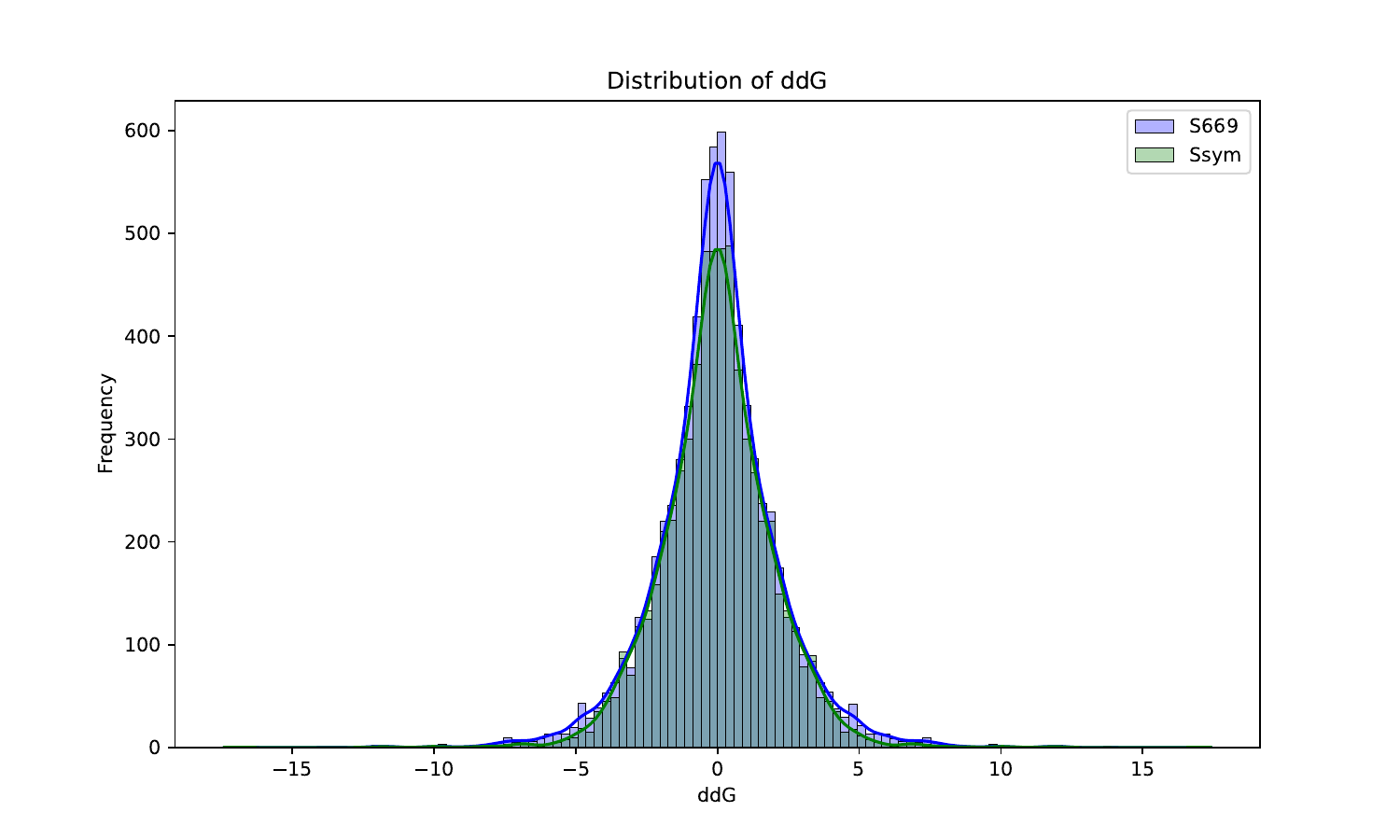}
\caption{Distribution of DDG in our proposed datasets} 
\label{fig:app:ddg_distri}
\end{figure}

\subsection{Sample Value Distribution}
\label{app:distribution}
From our curated dataset, we calculated the distribution of $\Delta \Delta G$ values, the results are visualized in Figure \ref{fig:app:ddg_distri}. It could be concluded that most values are centered around $[-5,5]$, with scarce exceptions.

\section{Implementation Details} 
\label{app:imple}
In the experiments, we trained the model for 25 epochs with a batch size of 64 and using the OneCycle optimizer with a learning rate of 0.001. The hyperparameters are obtained by conducting 5-fold cross-validation for best-averaged model performance in the validation datasets. The model was implemented based on the standard PyTorch Geometric~\cite{fey2019fast} library and Pytorch Lighting Framework. We ran the models on Intel(R) Xeon(R) Gold 6240R CPU @ 2.40GHz CPU and NVIDIA A100 GPU. 

\section{Performance Evaluation on S669 and Ssym Datasets}
\label{app:add_val}

Beyond the comparison and analysis in Section \ref{exp:eval}, we compared our model with other approaches aiming to predict the thermostability changes upon protein single-point mutation. Since the structures of some methods are complicated to re-implement into our framework to enable fair comparison, and some methods are only offered in the web server, we only tested the model performance on the S669 and Ssym datasets, without applying our training data to them or compare the computational efficiency. The results are shown in Table \ref{tab:app_eval_669} and Table \ref{tab:app_eval_sym}. Some metrics are taken from a comprehensive survey on predicting protein mutation effects \cite{pancotti2022predicting}. Methods in \textit{Italic} indicate structure-based methods, others are sequence-based ones.

\begin{table}[ht]
\vspace{-2mm}
\centering
\caption{The additional evaluation results on the S669 and S669\_r datasets. }
\label{tab:app_eval_669}
\setlength{\tabcolsep}{1.5mm}{
\begin{tabular}{c|ccc|ccc}
\toprule
Method & \multicolumn{3}{c|}{S669} & \multicolumn{3}{c}{S669\_r}  \\
\cmidrule{2-4} \cmidrule{5-7}
& PCC  $\uparrow$ & RMSE  $\downarrow$ & MAE $\downarrow$ & PCC $\uparrow$ & RMSE $\downarrow$ & MAE $\downarrow$ \\
\midrule
DDGun          & 0.41 & 1.72 & 1.25 & 0.38 & 1.75 & 1.24  \\
INPS-Seq          & 0.43 & 1.51 & 1.09 & 0.43 & 1.53 & 1.10 \\
\textit{DDGun3D}    & 0.43 & 1.6 & 1.11 & 0.41 & 1.62 & 1.14 \\
\textit{PremPS}    & 0.41 & 1.5 & 1.08 & 0.42 & 1.49 & 1.05 \\
\textit{MAESTRO}    & 0.5 & 1.44 & 1.06 & 0.2 & 2.1 & 1.66 \\
\textit{PoPMuSiC}    & 0.41 & 1.51 & 1.09 & 0.24 & 2.09 & 1.64 \\
\textit{ThermoNet}    & 0.39 & 1.62 & 1.17 & 0.38 & 1.66 & 1.23 \\
\hline
\rowcolor{gray!15} Ours            & 0.51 & 1.46 & 1.07 & 0.51 & 1.46 & 1.07 \\
\bottomrule
\end{tabular}}
\end{table}

\begin{table}[ht]
\vspace{-2mm}
\centering
\caption{The additional evaluation results on the Ssym and Ssym\_r datasets. }
\label{tab:app_eval_sym}
\setlength{\tabcolsep}{1.5mm}{
\begin{tabular}{c|ccc|ccc}
\toprule
Method & \multicolumn{3}{c|}{Ssym} & \multicolumn{3}{c}{Ssym\_r}  \\
\cmidrule{2-4} \cmidrule{5-7}
& PCC  $\uparrow$ & RMSE  $\downarrow$ & MAE $\downarrow$ & PCC $\uparrow$ & RMSE $\downarrow$ & MAE $\downarrow$ \\
\midrule
DDGun          & 0.51 & 1.51 & 1.12 & 0.51 & 1.51 & 1.12  \\
INPS-Seq          & 0.5 & 1.53 & 1.11 & 0.51 & 1.52 & 1.11 \\
\textit{DDGun3D}    & 0.58 & 1.45 & 1.04 & 0.56 & 1.49 & 1.07 \\
\textit{PremPS}    & 0.81 & 1.05 & 0.7 & 0.73 & 1.21 & 0.83 \\
\textit{MAESTRO}    & 0.6 & 1.37 & 0.95 & 0.25 & 2.27 & 1.73 \\
\textit{PoPMuSiC}    & 0.65 & 1.27 & 0.9 & 0.28 & 2.25 & 1.71 \\
\textit{ThermoNet}    & 0.45 & 1.66 & 1.16 & 0.39 & 1.73 & 1.21 \\
\hline
\rowcolor{gray!15} Ours             & 0.47 & 1.45 & 1.05 & 0.47 & 1.45 & 1.05 \\
\bottomrule
\end{tabular}}
\end{table}

It could be concluded from the results that our model demonstrates consistent performance across both datasets and their reversed counterparts. It is competitive in all metrics but is occasionally surpassed by models such as \textit{PremPS} \cite{chen2020premps} and \textit{MAESTRO} \cite{laimer2015maestro} in certain metrics. The strength of our model lies in its generalizability since there is no information leakage between the training dataset and the testing dataset while others exist or are unknown. Also, our model stands out for its consistent performance, while some other models exhibit more variability between datasets. Furthermore, numerous methods display comparable performance on these datasets. This underscores the importance of broadening the data spectrum and constructing more robust datasets for future research.

\end{document}